\pdfoutput=1

\newcommand{\caproman}[1]{\uppercase\expandafter{\romannumeral#1}}

\newcommand{\BGsite}{http://users.physics.harvard.edu/\raisebox{-4pt}{$\tilde{\;\;}$}gottschalk}

\newcommand{\BGmail}{bernardgottschalk\,@\,gmail.com}
\newcommand{\EWCmail}{ecascio\,@\,partners.org}

\newcommand{\fig}[1]{Fig.\,\ref{fig:#1}}

\documentclass[12pt]{article}

\title{\bf Using a 3D printer for 2D beam profile measurements in proton radiotherapy}
\author{Ethan W. Cascio\thanks{Francis H. Burr Proton Therapy Center, Massachusetts General Hospital, Boston, MA, USA, \EWCmail}~~and B. Gottschalk\thanks{Harvard University Laboratory for Particle Physics and Cosmology (LPPC), 18 Hammond St., Cambridge, MA 02138, USA, \BGmail}}
\date{\today}

\usepackage{amsmath}
\usepackage{latexsym}
\usepackage{graphicx}
\usepackage{rotating}
\usepackage{lineno}
\usepackage{setspace}
\usepackage[linktocpage=true,pdftex,hyperfigures=true]{hyperref}

\begin{document}

\maketitle


\begin{abstract}
\noindent We have used an inexpensive 3D printer mounting an inexpensive pre-irradiated silicon diode to measure the 2D dose distribution of a small ($\approx10$\,mm FWHM) proton beam. 

$z$ was measured with high resolution whereas $x$ was changed on a 2\,mm grid for a total of 12 scans and 2 repeats in 20 minutes. The peak dose to the diode was $\approx530$\,cGy/s, and no degradation in diode sensitivity was observed.

We present beam profiles and a 2D beam contour derived from these data. Most of the scan time was spent changing $x$ by hand, and a more appropriate 3D printer along with an improved setup and printer control software would reduce that to about 1 minute for an end-to-end absolute 2D dose measurement.

We propose extending this to 3D (2D transverse plus depth) for appropriate small beams by adding a small water tank.
\end{abstract}

\tableofcontents\clearpage

\section{Introduction}\label{sec:Introduction}
Options abound for 2D dose measurement in radiotherapy: ion chamber arrays, diode arrays, x-ray plates, film, and radiochromic film to name the more common. Even so, small beams, or larger beams with sharp features, favor scanning the field in 2D with a single small dosimeter. Computer-controlled x-y tables are available but many, designed for scientific applications, are overqualified in one respect (precision) and deficient in others (speed, range of motion, and cost).

By contrast, the $z$ (vertical) and $x$ (transverse) axes of typical 3D printers are well matched to our needs: accuracy of order $\pm0.1$\,mm or better, $15\times15$\,cm travel or greater, and high speed. Moreover, many are consumer products whose price has plummeted in recent years, to \$250 or less.

We have done a preliminary measurement in the Francis H. Burr Proton Therapy Center proton test beam. After describing the equipment, experimental procedure and analysis we suggest improvements that should bring the time down to $\approx1$\,minute for an end-to-end 2D absolute dose measurement.

\section{Equipment}\label{sec:Equipment}
\begin{figure}[p]
\centering\includegraphics[height=3.5in]{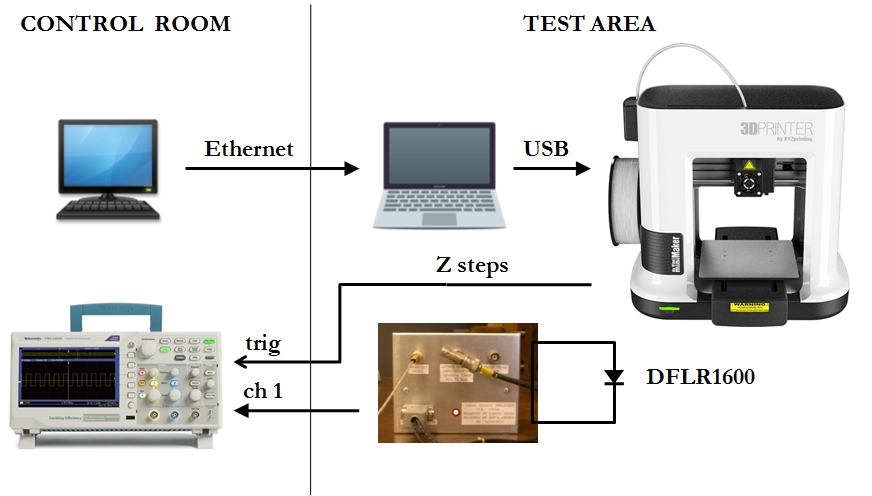}
\caption{Block diagram.}\label{fig:BlockDiagram}
\end{figure}

\fig{BlockDiagram} shows the setup. To respect USB cable length restrictions we connected the printer, a da Vinci MiniMaker by XYZprinting,\footnote{~Not recommended! See Discussion.} to a local laptop controlled via Ethernet. We tapped into the stepmotor pulses of the $z$ drive, and a 10\,s sweep of the LeCroy WavePro 715ZI oscilloscope was triggered by the first pulse. Previously, we had determined the $z$ scan speed to be a constant 7.28\,mm/s after a negligibly short acceleration time.

\begin{figure}[p]
\centering\includegraphics[height=3.5in]{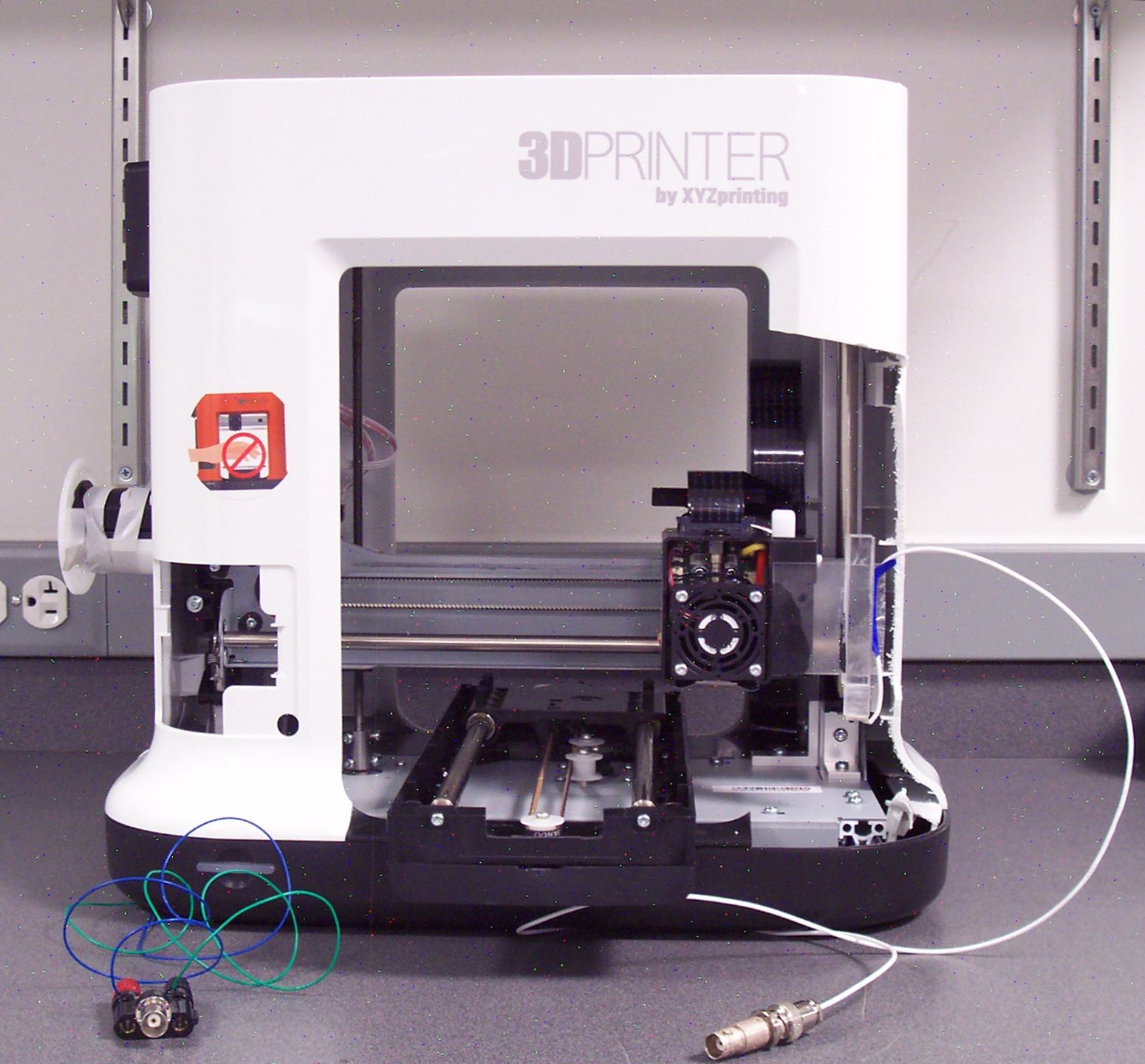}
\caption{3D printer with diode mounted at bottom of plastic bar to the right. The wires visible on the left carried the $z$ motor step pulses.}\label{fig:Scanner}
\end{figure}

A DFLR1600 diode was attached to the printer's nozzle carriage (\fig{Scanner}). This diode is a surface-mount 1A rectifier (\$0.46 in singles) whose properties as a proton dosimeter and/or fluence meter we have studied extensively
 \cite{Cascio2011}. The active area is $1\times1\times$\,mm$^2$ and the active thickness is that of the depletion layer. The diode was edge-on to the beam, presenting a horizontal 1\,mm line segment normal to the $z$ axis. Thus its size in $z$ was tiny. In $x$ it was 1\,mm with $x_\mathrm{rms}=0.5\,\mathrm{mm}/\sqrt{3}=0.29$\,mm. 
 
\begin{figure}[p]
\centering\includegraphics[height=3.5in]{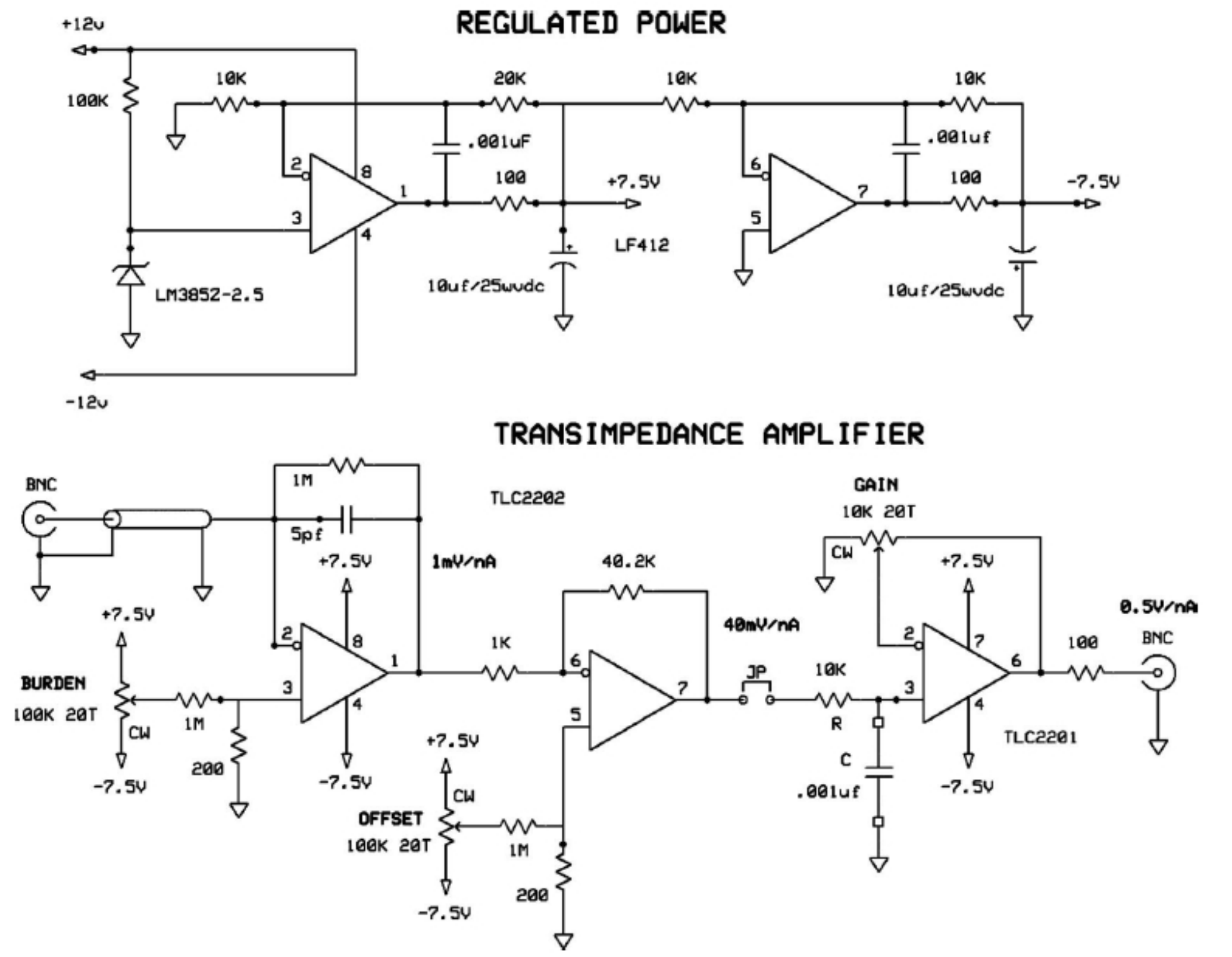}
\caption{Fast current-to-voltage amplifier used for these measurements. Contrary to the diagram, the gain was set to 0.103\,V/nA.}\label{fig:diodeAmplifier}
\end{figure}

The diode was pre-irradiated to 1\,KGy after which we found its sensitivity to be 0.107\,nC/(cGy to water). We used an amplifier of in-house design and construction (schematic diagram in \fig{diodeAmplifier}) whose gain (contrary to the diagram) was measured to be 0.103\,V/nA. The amplifier was reasonably fast ($\approx5\mu$\,s) so integration of the current was essentially done in software as will be described.

The beam stop was fairly far downstream and it is noteworthy that the laptop and 3D printer operated normally throughout, despite the room background (which we did not measure).

There was no beam monitor. We set the cyclotron for the lowest possible stable external current ($\approx1$\,nA), which proved adequate.

\section{Procedure}\label{sec:Procedure}

The proton beam energy was 228\,MeV. After rough alignment of the diode using radiochromic film, data acquisition proceeded as follows:
\begin{enumerate}
\item Turn beam on.
\item Trim beam current to 1\,nA using in-house ratemeter.
\item Scan $z$ (10\,s = 72.8\,mm).
\item Turn beam off.
\item Name and save data file. The file name encoded $x$.
\item Jog $x$ (multiple of 2\,mm).
\item Zero $z$.
\item Repeat.
\end{enumerate}
We took scans at 12 interleaved values of $x$, followed by two repeat scans. The entire experiment took 20\,minutes, most of it spent preparing for the next scan.

\section{Analysis}\label{sec:Analysis}
\begin{figure}[p]
\centering\includegraphics[height=3.5in]{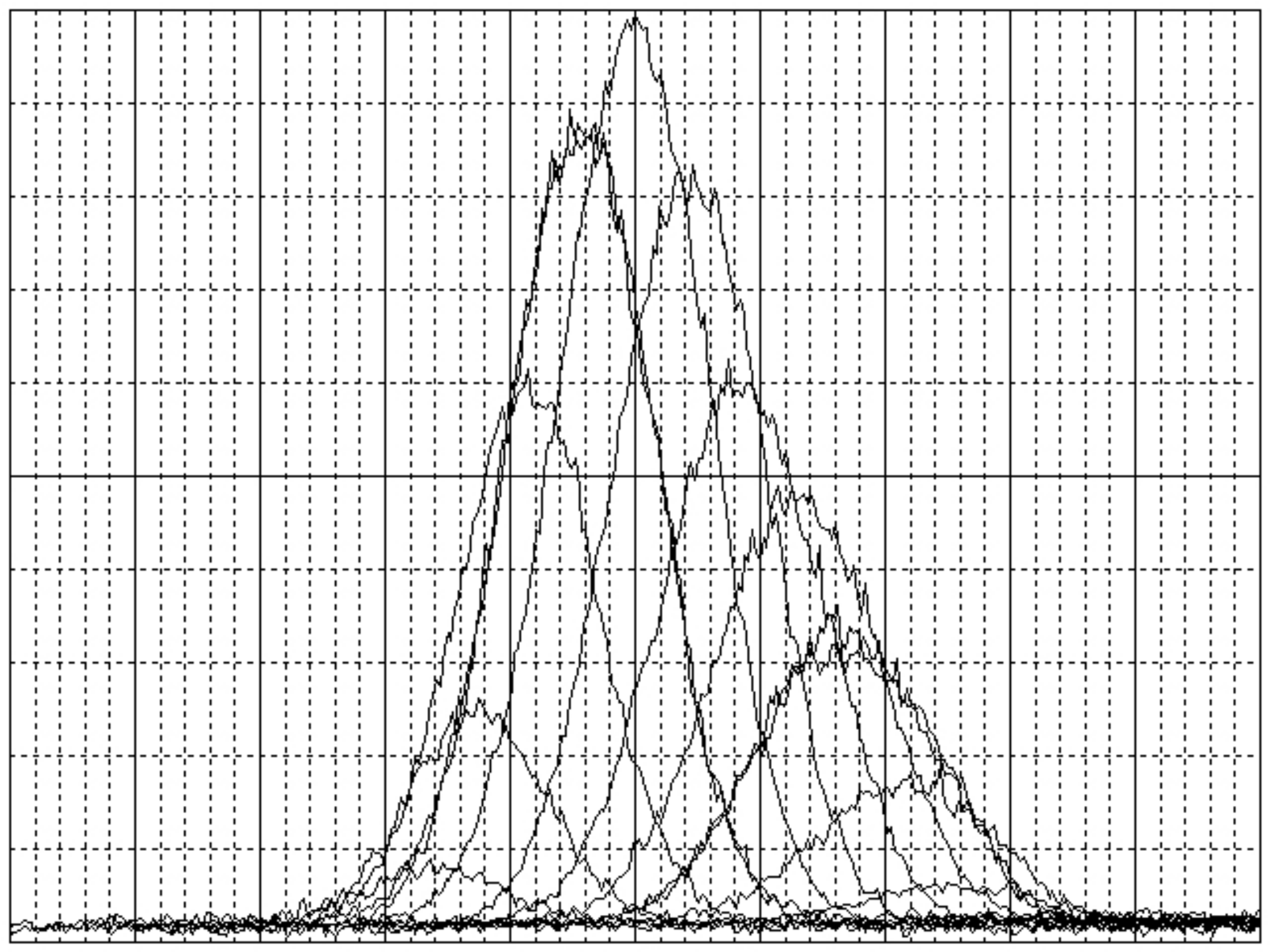}
\caption{Raw scans in $z$ (1\,mm/minor division), each offset by its $x$ value (mm).}\label{fig:rawScan}
\end{figure}

Ideally, an analysis program would be ready to go. In this instance, the analysis was done after the fact, off-site and off-line. The Lecroy scope had inadvertently been set so the 14 $z$-scan text files had 25\,K lines each, so they were immediately averaged by fifties to obtain 500-line text files, which we will call `raw data', each point representing 20\,ms. \fig{rawScan} shows these scans with background subtracted and each scan offset by its nominal $x$ value. 

\begin{figure}[p]
\centering\includegraphics[height=3.5in]{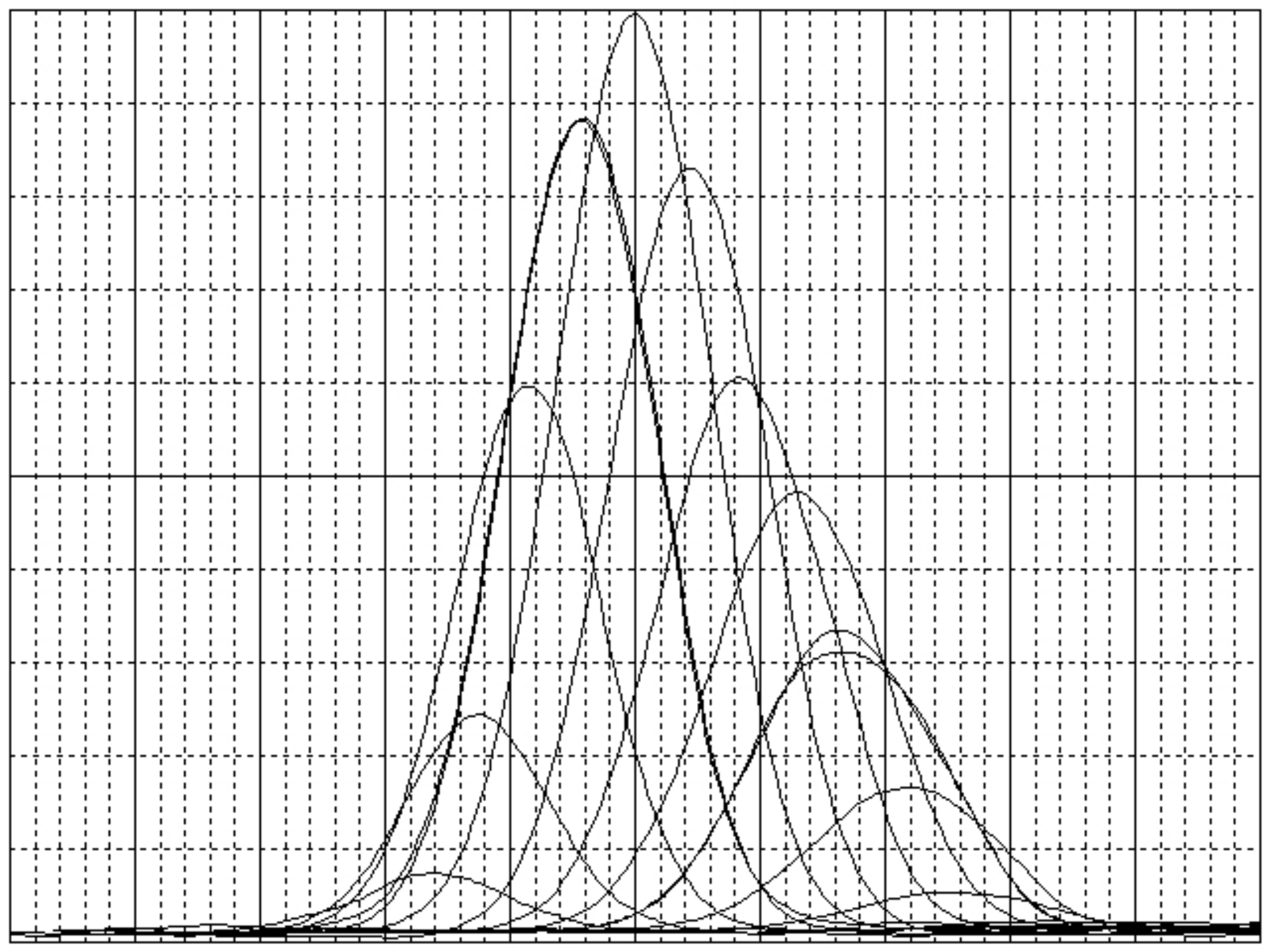}
\caption{Smoothed scans in $z$ (1\,mm/minor division), each offset by its $x$ value (mm). The peak signal is 5.67\,V.}\label{fig:smoothScan}
\end{figure}

The noise is not amplifier noise. It stems from the randomness of dose deposition and the very small active volume of the dosimeter, causing the total dose in each 20\,ms interval to vary slightly even at constant nominal dose. We therefore smooth the raw data by convolution with a normal 1D Gaussian:
\begin{equation}
V'_i\,=\,\sum_{j\,=\,i-5\sigma}^{i+5\sigma}V_j\;G(i-j;\sigma)
\end{equation}
where 
\begin{equation}\label{G1D}
G_\mathrm{1D}(k;\sigma)\;\equiv\;\frac{1}{\sqrt{2\pi}\,\sigma}\;e^{\displaystyle{-\frac{1}{2}\left(\frac{k}{\sigma}\right)^2}}
\end{equation}
and $\sigma$ (a tunable parameter) equals 5 raw data intervals (100\,ms). \fig{smoothScan} shows the smoothed data. Note the good agreement between the original and repeat runs showing, most likely, that change in diode sensitivity and variation in beam current during the experiment were negligible. 

\begin{figure}[p]
\centering\includegraphics[height=3.5in]{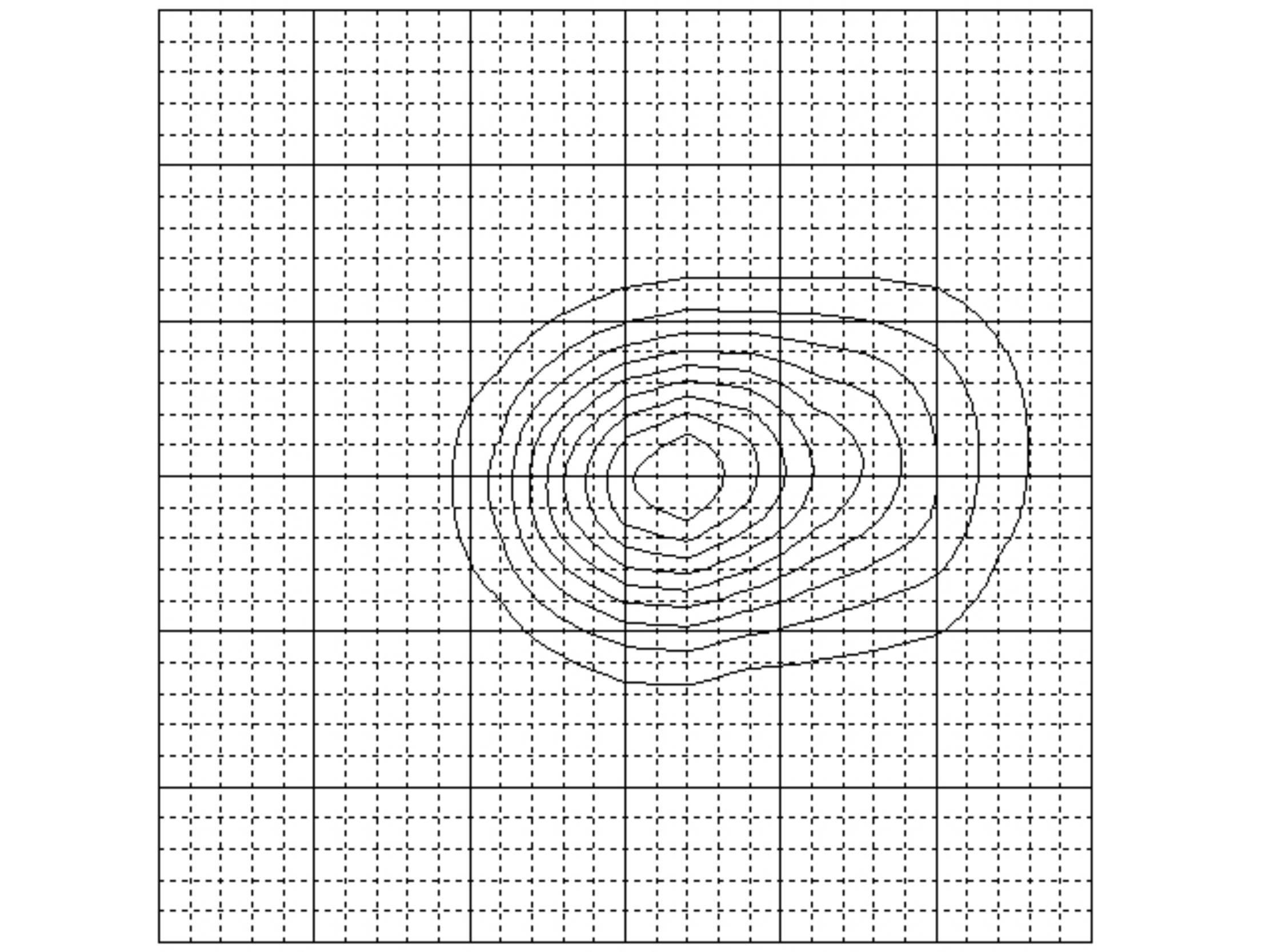}
\caption{Dose contours from 90\% to 10\% by 10\% steps, 1\,mm per minor division.}\label{fig:contours}
\end{figure}

Finally, \fig{contours} shows a contour plot of the first twelve scans. We used the Fortran contouring algorithm of Aramini \cite{Aramini1980}. The smaller spread in $z$ than in $x$ (the bend plane of the beam transport), as well as the tail in $x$, are frequent features of small unscattered proton beams.

The peak signal is consistent with the beam current, diode calibration, amplifier gain and contour plot, as we show now. Assume the fluence in the measuring plane is approximated by a cylindrical 2D Gaussian
\begin{equation}\label{eqn:G2D}
G_\mathrm{2D}(r,\phi;\sigma)\;\equiv\;\frac{1}{\pi\,\sigma_r^2}\;e^{\displaystyle{-\left(\frac{r}{\sigma_r}\right)^2}}
\end{equation}
whose effective area is $\pi\sigma_r^2$ with $\sigma_r=\sqrt{2}\,r_{60}$ where $r_{60}$ is the radius at the 60\% point, the fourth contour. The beam is obviously {\em not} circular but we compensate for that by using $r_{60}=3.5$\,mm, the mean  between $x$ and $z$ ignoring the $x$ tail. That yields a beam effective area of 0.77\,cm$^2$. The dose rate in practical units \cite{BGch2v2arXiv} is
\begin{equation}\label{eqn:Ddot}
\dot{D}\;=\;\frac{i_\mathrm{P}}{A}\;\left(\frac{S}{\rho}\right)\qquad\mathrm{Gy/s}
\end{equation}
where $i_\mathrm{P}$ is the proton current in nA, $A$ is the effective area in cm$^2$ and $S/\rho$ is the mass stopping power in the `dose to' material in MeV/(g/cm$^2$). Using Janni \cite{janni82} as implemented in LOOKUP \cite{lookup} for the stopping power in water, and plugging in the numbers, we find
\begin{equation}\label{eqn:Vpeak}
V_\mathrm{peak,\,predicted}\;=\;\frac{4.105}{0.77}\;\frac{\mathrm{Gy}}{\mathrm{s}}\times10.7\;\frac{\mathrm{nC}}{\mathrm{Gy}}\times0.103\;\frac{\mathrm{V}}{\mathrm{nA}}\;=\;5.88\,\mathrm{V}
\end{equation}
in satisfactory agreement with $V_\mathrm{peak,\,measured}=5.67$\,V.

Analysis time is negligible. It took 4\,s on a Lenovo T400 laptop running Intel Fortran to process 14 scans and prepare the graphs.

\section{Discussion}\label{sec:Discussion}
We are improving the equipment and procedure as follows:
\begin{enumerate}
\item Obtain a 3D printer that accepts G code and Repetier open-source software so as to program a raster scan, with repeats, launched by a single keystroke.
\item Active USB extender or Ethernet-capable 3D printer to simplify setup.
\item Reduce $z$ overscan.
\item Record $x$ and $z$ stepmotor pulses (scope channels C2,  C3) to eliminate reliance on constant speed.
\item Record beam monitor ionization chamber (C4) to eliminate reliance on constant beam current.
\end{enumerate}
With those improvements, and assuming an analysis program ready to go, results similar to those presented could be obtained in about a minute.

Of course, that does not take full advantage of 3D printer capabilities. One can imagine putting a small water tank on the table and using it to do a full 3D beam scan, or transverse scans for centering followed by a scan in depth.

Pending these improvements and extensions, we have already shown that a 3D printer combined with a diode is a viable option for the rapid characterization of small beams.

\end{document}